


\font\bigbold=cmbx12
\font\eightrm=cmr8
\font\sixrm=cmr6
\font\fiverm=cmr5
\font\eightbf=cmbx8
\font\sixbf=cmbx6
\font\fivebf=cmbx5
\font\eighti=cmmi8  \skewchar\eighti='177
\font\sixi=cmmi6    \skewchar\sixi='177
\font\fivei=cmmi5
\font\eightsy=cmsy8 \skewchar\eightsy='60
\font\sixsy=cmsy6   \skewchar\sixsy='60
\font\fivesy=cmsy5
\font\eightit=cmti8
\font\eightsl=cmsl8
\font\eighttt=cmtt8
\font\tenfrak=eufm10
\font\sevenfrak=eufm7
\font\fivefrak=eufm5
\font\tenbb=msbm10
\font\sevenbb=msbm7
\font\fivebb=msbm5
\font\tensmc=cmcsc10
\font\tencmmib=cmmib10  \skewchar\tencmmib='177
\font\sevencmmib=cmmib10 at 7pt \skewchar\sevencmmib='177
\font\fivecmmib=cmmib10 at 5pt \skewchar\fivecmmib='177

\newfam\bbfam
\textfont\bbfam=\tenbb
\scriptfont\bbfam=\sevenbb
\scriptscriptfont\bbfam=\fivebb

\newfam\frakfam
\textfont\frakfam=\tenfrak
\scriptfont\frakfam=\sevenfrak
\scriptscriptfont\frakfam=\fivefrak

\newfam\cmmibfam
\textfont\cmmibfam=\tencmmib
\scriptfont\cmmibfam=\sevencmmib
\scriptscriptfont\cmmibfam=\fivecmmib
\def\bold#1{\fam\cmmibfam\relax#1}


\def\eightpoint{%
\textfont0=\eightrm   \scriptfont0=\sixrm
\scriptscriptfont0=\fiverm  \def\rm{\fam0\eightrm}%
\textfont1=\eighti   \scriptfont1=\sixi
\scriptscriptfont1=\fivei  \def\oldstyle{\fam1\eighti}%
\textfont2=\eightsy   \scriptfont2=\sixsy
\scriptscriptfont2=\fivesy
\textfont\itfam=\eightit  \def\it{\fam\itfam\eightit}%
\textfont\slfam=\eightsl  \def\sl{\fam\slfam\eightsl}%
\textfont\ttfam=\eighttt  \def\tt{\fam\ttfam\eighttt}%
\textfont\bffam=\eightbf   \scriptfont\bffam=\sixbf
\scriptscriptfont\bffam=\fivebf  \def\bf{\fam\bffam\eightbf}%
\abovedisplayskip=9pt plus 2pt minus 6pt
\belowdisplayskip=\abovedisplayskip
\abovedisplayshortskip=0pt plus 2pt
\belowdisplayshortskip=5pt plus2pt minus 3pt
\smallskipamount=2pt plus 1pt minus 1pt
\medskipamount=4pt plus 2pt minus 2pt
\bigskipamount=9pt plus4pt minus 4pt
\setbox\strutbox=\hbox{\vrule height 7pt depth 2pt width 0pt}%
\normalbaselineskip=9pt \normalbaselines
\rm}


\def\pagewidth#1{\hsize= #1}
\def\pageheight#1{\vsize= #1}
\def\hcorrection#1{\advance\hoffset by #1}
\def\vcorrection#1{\advance\voffset by #1}

\newcount\notenumber  \notenumber=1              
\newif\iftitlepage   \titlepagetrue              
\newtoks\titlepagefoot     \titlepagefoot={\hfil}
\newtoks\otherpagesfoot    \otherpagesfoot={\hfil\tenrm\folio\hfil}
\footline={\iftitlepage\the\titlepagefoot\global\titlepagefalse
           \else\the\otherpagesfoot\fi}

\def\abstract#1{{\parindent=30pt\narrower\noindent\eightpoint\openup
2pt #1\par}}
\def\smc{\tensmc}


\def\note#1{\unskip\footnote{$^{\the\notenumber}$}
{\eightpoint\openup 1pt
#1}\global\advance\notenumber by 1}

\def\frac#1#2{{#1\over#2}}
\def\dfrac#1#2{{\displaystyle{#1\over#2}}}

\def\({\left(}
\def\){\right)}
\def\<{\langle}
\def\>{\rangle}
\def\2pd#1#2#3{\frac{\partial^2#1}{\partial#2\partial#3}}

\def\sqr#1#2{{\vcenter{\vbox{\hrule height.#2pt
        \hbox{\vrule width.#2pt height#1pt \kern#1pt
           \vrule width.#2pt}
        \hrule height.#2pt}}}}

\def\ni{\noindent}
\def\lqq{\lq\lq}
\def\rqq{\rq\rq}
\def\slash{\!\!\!/\,}


\global\newcount\secno \global\secno=0
\global\newcount\meqno \global\meqno=1
\global\newcount\appno \global\appno=0
\newwrite\eqmac
\def\romappno{\ifcase\appno\or A\or B\or C\or D\or E\or F\or G\or H
\or I\or J\or K\or L\or M\or N\or O\or P\or Q\or R\or S\or T\or U\or
V\or W\or X\or Y\or Z\fi}
\def\eqn#1{
        \ifnum\secno>0
            \eqno(\the\secno.\the\meqno)\xdef#1{\the\secno.\the\meqno}
          \else\ifnum\appno>0
            \eqno({\rm\romappno}.\the\meqno)\xdef#1{{\rm\romappno}.
               \the\meqno}
          \else
            \eqno(\the\meqno)\xdef#1{\the\meqno}
          \fi
        \fi
\global\advance\meqno by1 }

\def\eqnn#1{
        \ifnum\secno>0
            (\the\secno.\the\meqno)\xdef#1{\the\secno.\the\meqno}
          \else\ifnum\appno>0
            \eqno({\rm\romappno}.\the\meqno)\xdef#1{{\rm\romappno}.
                \the\meqno}
          \else
            (\the\meqno)\xdef#1{\the\meqno}
          \fi
        \fi
\global\advance\meqno by1 }

\global\newcount\refno
\global\refno=1 \newwrite\reffile
\newwrite\refmac
\newlinechar=`\^^J
\def\ref#1#2{\the\refno\nref#1{#2}}
\def\nref#1#2{\xdef#1{\the\refno}
\ifnum\refno=1\immediate\openout\reffile=refs.tmp\fi
\immediate\write\reffile{
     \noexpand\item{[\noexpand#1]\ }#2\noexpand\nobreak.}
     \immediate\write\refmac{\def\noexpand#1{\the\refno}}
   \global\advance\refno by1}
\def\semi{;\hfil\noexpand\break ^^J}
\def\nl{\hfil\noexpand\break ^^J}
\def\refn#1#2{\nref#1{#2}}

\def\up#1{$^{[#1]}$}

\def\cmp#1#2#3{{\it Commun. Math. Phys.} {\bf {#1}} (19{#2}) #3}
\def\jmp#1#2#3{{\it J. Math. Phys.} {\bf {#1}} (19{#2}) #3}

\def\pl#1#2#3{{\it Phys. Lett.} {\bf {#1}B} (19{#2}) #3}
\def\np#1#2#3{{\it Nucl. Phys.} {\bf B{#1}} (19{#2}) #3}

\def\pr#1#2#3{{\it Phys. Rev.} {\bf {#1}} (19{#2}) #3}

\def\prD#1#2#3{{\it Phys. Rev.} {\bf D{#1}} (19{#2}) #3}
\def\prl#1#2#3{{\it Phys. Rev. Lett.} {\bf #1} (19{#2}) #3}

\def\ann#1#2#3{{\it Ann. Phys.} {\bf {#1}} (19{#2}) #3}

\def\zpC#1#2#3{{\it Z. Phys.} {\bf C{#1}} (19{#2}) #3}

\def\fortschr#1#2#3{{\it Fortschr. d. Phys.} {\bf {#1}} (19{#2}) #3}

\def\cjp#1#2#3{{\it Can. J. Phys.} {\bf #1} (19{#2}) #3}

\def\empty#1#2#3{{\bf{#1}} (19{#2}) #3}


\def\a{\alpha}
\def\b{\beta}

\def\eps{\epsilon}

\def\psic{\psi_{\rm c}}
\def\psiv{\psi_{\bold v}}
\def\d{\delta}

\def\pa{\partial}
\def\I{{\bold I}}
\def\vdenom#1#2{{\left(\dfrac{(#1_1-#2_1)^2}{1-v^2}+ (#1_2-#2_2)^2+
(#1_3-#1_3)^2\right)}}


\pageheight{24cm}
\pagewidth{15.5cm}
\magnification \magstep1
\voffset=8truemm
\baselineskip=16pt
\parskip=5pt plus 1pt minus 1pt


\input epsfig.sty
\def\fig#1#2{\vbox{\epsfig{file=#1,width=#2}}}

\secno=0

{\eightpoint
\refn\JAUCH{J.M.\ Jauch and F.\ Rohrlich, {\it The Theory
of Photons and Electrons}, Second Expanded
Edition, (Springer-Verlag, New York 1980)}
\refn\FERMI{E. Fermi, {\it Atti della Reale Accademia dei Lincei}, {\bf
12} {(1930)} {431}}
\refn\STROCCHI{F.\ Strocchi and A.S.\ Wightman, \jmp{15}{74}{2198}}
\refn\ZWANZIG{D.\ Maison and D.\ Zwanziger, \np{91}{75}{425}}
\refn\MONSTER{M. Lavelle and D. McMullan, {\sl Constituent Quarks from
QCD},\nl Barcelona/Plymouth preprint UAB-FT-369/PLY-MS-95-03}
\refn\HAAG{R.\ Haag, {\sl Local Quantum Physics}, (Springer-Verlag,
Berlin, Heidelberg, 1993)}
\refn\BUCHHOLZ{D.\ Buchholz, \pl{174}{86}{331}}
\refn\DIRACINCANADA{P.A.M.\ Dirac, \cjp{33}{55}{650}}
\refn\MANDEL{S.\ Mandelstam, \ann{19}{62}{1}}
\refn\STEINMANN{O.\ Steinmann, \ann{157}{84}{232}}
\refn\STEINMANNAGAIN{O.\ Steinmann, \np{350}{91}{355};
\empty{361}{91}{173}}
\refn\EMILIOE{E.\ d'Emilio and M.\ Mintchev, \fortschr{32}{84}{473},
503}
\refn\BUCHSTRING{D.\ Buchholz, \cmp{85}{82}{49}}
\refn\SYMM{M.\ Lavelle and D.\ McMullan, \prl{71}{93}{3758}}
\refn\PANP{M. Lavelle and D. McMullan, \pl{329}{94}{68}}
\refn\SSB{M.\ Lavelle and D.\ McMullan, \pl{347}{95}{89}}
\refn\COLOUR{M.\ Lavelle and D.\ McMullan, {\sl The Colour of Quarks},
\nl to appear in {\sl Phys.\ Lett.\ \bf B}.}
\refn\DIRAC{P.A.M. Dirac, \lqq Principles of Quantum Mechanics\rqq,
(OUP, Oxford, 1958), page 302}
\refn\TOM{J.C.\ Breckinridge, M.\ Lavelle and T.G.\ Steele,
\zpC{65}{95}{155}}
\refn\ADKINS{G.S.\ Adkins, \prD{27}{83}{1814}}
\refn\HECK{D.\ Heckathorn, \np{156}{79}{328}}
\refn\DMONE{R.\ Tarrach, \np{183}{81}{384}}
\refn\BROADHURST{N.\ Gray, D.J.\ Broadhurst, W.\ Grafe and
K.\ Schilcher, \zpC{48}{90}{673}}
\refn\YENNIE{H.M.\ Fried and D.R.\ Yennie, \pr{112}{58}{1391}}
\refn\NAKA{N.\ Nakanishi and I.\ Ojima, {\sl Covariant Operator
Formalism of Gauge Theories and Quantum Gravity}, (World Scientific,
Singapore 1990)}
\refn\AXIAL{See, e.g., {\sl Physical and Non-Standard Gauges}, ed.'s
P.\ Gaigg et al, Springer Lecture Notes in Physics {\bf 361}
(Springer-Verlag, Heidelberg 1990)}
}
%
%
\rightline {UAB-FT-379}
\rightline {PLY-MS-95-08}
\vskip 16pt
\centerline{\bigbold THE PHYSICAL PROPAGATOR OF A}
\centerline{\bigbold SLOWLY MOVING CHARGE}
\vskip 20pt
\centerline{\smc Emili Bagan{\hbox
{$^1$}}\footnote{{*}}{{\eightpoint\rm Permanent
address and after 1.1.1996:
IFAE, Universitat Aut\`onoma de Barcelona.}},
Martin Lavelle{\hbox {$^2$}}
and  David McMullan{\hbox {$^3$}}}
\vskip 15pt
{\baselineskip 12pt
\centerline{\null$^1$Physics Department}
\centerline{Bldg.\ 510A}
\centerline{Brookhaven National Laboratory}
\centerline{Upton, NY 11973}
\centerline{USA}
\centerline{email: iftebag@cc.uab.es}
\vskip 13pt
\centerline{\null$^2$Grup de F\'\i sica Te\`orica and IFAE}
\centerline{Edificio Cn}
\centerline{Universitat Aut\`onoma de Barcelona}
\centerline{E-08193 Bellaterra (Barcelona)}
\centerline{Spain}
\centerline{email: lavelle@ifae.es}
\vskip 13pt
\centerline{\null$^{3}$School of Mathematics and Statistics}
\centerline{University of Plymouth}
\centerline{Drake Circus, Plymouth, Devon PL4 8AA}
\centerline{U.K.}
\centerline{email: d.mcmullan@plymouth.ac.uk}}
\vskip 1truemm
\vskip 12pt
{\baselineskip=13pt\parindent=0.58in\narrower\ni{\bf Abstract}\hskip
2truemm
We consider an electron which is electromagnetically dressed in such a
way that it is both gauge invariant and that
it has the associated electric and magnetic fields expected of a moving
charge. We study the propagator of this dressed electron
and, for small velocities, show explicitly at one loop that at the
natural (on-shell) renormalisation point, $p_0=m, {\bold p}=m
{\bold v}$, one can renormalise the propagator multiplicatively.
Furthermore the renormalisation constants are infra-red finite.
This shows that the dressing we use corresponds
to a slowly moving, physical asymptotic field.
\par}

\vfill\eject
\noindent It is well known that there are difficulties in finding the
correct asymptotic fields in theories with massless particles
and/or  confinement. These problems plague the gauge theories of the
electromagnetic and strong interactions. They find expression in
infra-red divergences in the perturbative expansion of Green's
functions with external lines corresponding to charged particles.
The standard approaches to the infra-red problem
dress the charged particles by adding an (infinite) number of the
massless gauge bosons. This means that descriptions of charged
particles should be based around coherent states and not on a Fock
space description. (For a concise review of the infra-red problem and a
critique of standard responses see Supplement 4 of Ref.\ \JAUCH.)
We also recall here that it is known that any description of a
state corresponding to a charged particle must be both
non-local\up{\FERMI-\MONSTER} and
non-covariant\up{\HAAG, \BUCHHOLZ, \MONSTER}.
Physically this merely corresponds to the need to
dress an electron, say, with an electromagnetic cloud which essentially
falls off away from the charge as $1/R$ and whose exact form depends
upon the velocity of the charge at its centre. Although the need for
such a dressing can often be sidestepped in the calculation of QED
scattering processes, it is fundamental to any description of charged
states in this theory. In QCD where charges are confined by their
mutual interactions, one finds\up{\MONSTER} that the dressings
play an even greater role, and, e.g., they are required if we want
to construct colour charges in a well-defined way.

There is a long history of trying to understand how dressings
can be incorporated into gauge
theories\up{\DIRACINCANADA-\BUCHSTRING}.
In a recent series of papers\up{\SYMM-\COLOUR}, which are summarised
and extended in Ref.\ \MONSTER,  two of us have
proposed a new approach to gauge theories where the dressing associated
with charged states is taken fully into account from the beginning. Two
criteria lie at the heart of this approach: the dressed
charged particle, since it describes a physical field, should be
gauge invariant and secondly
the associated electric and magnetic fields
should be those we expect of a charged particle. Such a dressed field
was, in fact, suggested many years ago by Dirac (see Sect.\ 80 of
Ref.\ \DIRAC). He pointed out that the gauge invariant combination
$$
\psic(x) =\exp\(ie\frac{\pa_iA_i}{\nabla^2}(x)\right)\psi(x)
\,,
\eqn\static
$$
made out of a fermion and a non-local cloud of vector potentials could
represent an electron since it generates a Coulomb electric field that
we would like to associate with a physical electron. To see this,
one may make use of the equal time
commutator, $[E_i({\bold x}),A_j({\bold y})]=
i\d_{ij}\d({\bold x}-{\bold y})$, and see that
for an eigenstate, $\vert \epsilon\rangle$ of the electric field
with eigenvalue, $\epsilon_i({\bold x})$, the introduction of such a
dressed charge as (\static) alters the electric field in the way we
expect of a physical static charge:
$$
E_i({\bold x})\psic({\bold y})\vert\epsilon\rangle
=
\( \epsilon_i({\bold x}) +\frac e{4\pi}\frac{{\bold x}_i-{\bold
y}_i}{\vert{\bold x-y}\vert^3}
\)\psic({\bold y})\vert\epsilon\rangle
\,.
\eqn\statcomm
$$

It is evident that the description (\static) of an electron is, as one
expects, non-local and non-covariant. It describes
a static charge (one can
easily see for example that (\static) does not possess any
associated magnetic field) and should only be used for this purpose.
In a recent paper\up{\MONSTER}
a description of a dressed
abelian charge, moving with a velocity
${\bold v}$,
analogous to (\static) was found. For a charge moving in the $x^1$-direction,
so that ${\bold v}=(v,0,0)$
this is
$$
\eqalign{
\psi_{{\bold v}}=& \exp\Bigg(\Bigg.
-\frac{e}{4\pi}\frac1{\sqrt{1-v^2}}\cr
&\times\int d^3z\frac{(1-v^2)\pa_1A_1(x^0,{\bold z})+\pa_2A_2(x^0,{\bold
z})+\pa_3A_3(x^0,{\bold z})-vE_1(x^0,{\bold
z})}{\vdenom{x}{z}^{\!\!\frac12}}\Bigg.\Bigg) \psi(x)\,,
}
\eqn\nonstatic
$$
which is easily seen to be gauge invariant and to reduce to (\static)
in the limit, ${\bold v}\to 0$. From the fundamental commutators one
can straightforwardly find that the magnetic and electric fields
associated with (\nonstatic) are just those that we would expect from a
charge moving with such a velocity ${\bold v}$.

One might wonder if this argument, based on the use of free
commutators, really holds in the interacting theory.
In this letter we will calculate the one loop propagator for such a
non-static
dressed charge and show that this interpretation of the dressed field
indeed persists in the quantum theory.
For computational simplicity we will restrict ourselves
to small velocities and work at order ${\bold v}$.
This means that we will find the one loop
propagator of
$$
\psi_{\bold v}(x)=
\exp\left(
ie\frac{\pa_jA_j+ v_i E_i}{\nabla^2}
\right)\psi(x)
\,.
\eqn\smallv
$$
Work on the, much more involved, relativistic case is in progress.
Note further
that $e^2$ terms in (\smallv) may be dropped since in the one
loop propagator they will only yield tadpoles which can be dropped in
BRST respecting regularisation schemes.

Before proceeding to the calculation of the propagator itself we should
briefly recall what is known about the two-point Green's function
of the Lagrangian
fermion at one loop and also the equivalent propagator for the dressed,
static charge, (\static). The renormalisation of this
Green's function of the Lagrangian field requires a mass
renormalisation and a wave function renormalisation. Working in
a general covariant gauge, one finds that the mass renormalisation is
gauge parameter independent, which indicates its physical significance
(see also Ref.\ \TOM). The wave function renormalisation
displays the infra-red singularities mentioned in the introduction if
one tries to renormalise this propagator on shell. This
shows that the Lagrangian fermion is not a good asymptotic field and
the need for a dressing.

The one loop propagator of the dressed static
fermion, (\static), was carried
out in Ref.\ \MONSTER. It was found that the propagator was gauge
independent (the calculation was carried out both in a general Lorentz
gauge and in Coulomb gauge) and that the mass renormalisation
required was just the standard one for the usual fermion propagator.
An on-shell  wave function renormalisation  was performed for the
static mass shell scheme (i.e., $p_0=m, {\bold p}=0$) whose use is, as
discussed above, a
natural consequence of (\statcomm) and
no difficulties were encountered. This was interpreted as strong
evidence that (\static) is a good asymptotic  description of a static
charge. It was further checked that (\static) is {\it not\/} suitable
for describing moving charges: any attempt at renormalising the
propagator of (\static) at a non-static mass shell was shown to
founder upon non-multiplicative structures and infra-red divergences.
This led to the conjecture in Ref.\ \MONSTER\ that it would be possible
to renormalise the propagator of (\nonstatic) at the relevant
non-static mass shell. After having
set the scene we may now proceed to the calculation of the one
loop propagator of (\smallv) and verify the above conjecture.

There are two ways to calculate this propagator. Either one may work in
what we call the {\sl dressing gauge}, where the dressing
vanishes\note{In general for three-point functions and higher there is
no gauge where all the dressings would be expected to vanish. Note also
that if we have more than one charged particle, there is a phase factor
as well as the various individual
dressings\up{\MONSTER}.}\ or one
can use a general Lorentz gauge. We have followed both approaches and
checked that the same results emerge. The dressing gauge here is
$\pa_i A_i+v_i E_i=0$ and one may readily see that the photon
propagator in this gauge is
$$
\eqalign{
D_{\mu\nu}(k)=\frac1{k^2}\Big(\Big.
-g_{\mu\nu}-& \frac{({\bold k}^2 -2k_0
v\cdot k )k_\mu k_\nu}{{\bold k}^4} \cr
& -\frac{k_0-k\cdot v}{-{\bold k}^2}
(k_\mu\eta_\nu+\eta_\mu k_\nu)
 -\frac{k_0}{-{\bold k}^2}(k_\mu v_\nu +
v_\mu k_\nu)\Big.\Big)
+O({\bold v}^2)
\,,}
\eqn\no
$$
where we have introduced the temporal vector,
$\eta=(1,0,0,0)$ and we point out that $v^\mu=(0, {\bold v})$ implies
that $v\cdot k=-{\bold v}\cdot{\bold k}$.
Working in this gauge one merely needs to
calculate the diagram of Fig.\ 1.a. In general we have contributions to
order $e^2$ from both interaction vertices and the expansion of the
dressing in (\smallv) and in an arbitrary Lorentz gauge all of
the diagrams of Fig.\ 1 must be calculated. We stress that the gauge
parameter dependence then cancels and that the result is the same as
that found in our dressing gauge.
\medskip
\midinsert
\fig{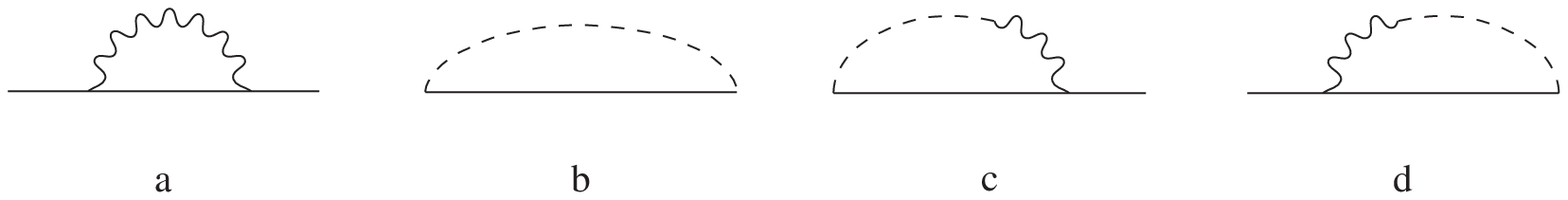}{12cm}
\smallskip{\eightpoint\narrower\noindent{\bf Figure 1}
The diagrams which yield the one loop dressed propagator. In the
appropriate dressing gauge only Fig.\ 1.a contributes. In a general
gauge all the diagrams must be evaluated. The dashed lines indicate
the projection of the photon propagator from the
(${\bold v}$-dependent) vertices in the dressing.\par}
\endinsert

Since we work to order ${\bold v}$ our propagator is a sum of two
terms: those which one has in Coulomb gauge and various new terms
linear in the velocity. The former terms have been considered in
Ref.'s \ADKINS\ (where a list
of integrals may be found) and \MONSTER. The latter terms yield
some new ultraviolet divergences. Dropping, for the
moment, the finite terms we find with a little effort the
following UV divergences which are proportional to ${\bold v}$ in the
self-energy:
$$
-i\Sigma^{\rm UV}_{\bold v}=\frac{i\a}{4\pi}\frac1{2-\omega}\(
\frac83p_0 v\slash- \frac83 p\cdot v\eta\slash
\) +\cdots
\,,
\eqn\UVs
$$
and $\alpha=e^2/4\pi$.
Since these tensor structures do not appear in the original Lagrangian
we would seem not to be able to perform a multiplicative
renormalisation. Although this difficulty is common in axial
gauge studies, its appearance would be unattractive. Recall also
that
in Coulomb gauge it has been shown by Heckathorn\up{\HECK} that the UV
divergences in the propagator are multiplicatively renormalisable.
We will see, however,
that these divergences can in fact be dealt with straightforwardly.
It should also perhaps be noted that this
apparent difficulty is not present on-shell. If we use
$\bar\psi\gamma_\mu\psi=\bar\psi\psi p_\mu/m$, which
holds for on-shell spinors, then it immediately follows that the UV
divergences of (\UVs) vanish on-shell.

To renormalise the propagator through mass and
wave-function counterterms in a general and systematic way
consider now the following (matrix)
multiplicative
renormalisation scheme:
$$
\psiv^B= \Big( Z_2^{\frac12} \I +\frac{Z'}{Z_2^{\frac12}}\eta\slash
v\slash \Big) \psiv\,,
\eqn\theZs
$$
or equivalently  at lowest order
$$
\psiv^B
= \sqrt{Z_2}\exp\{ -i\frac{Z'}{Z_2}\sigma^{\mu\nu}\eta_\mu v_\nu\}
\psiv
\,,
\eqn\theZsmatrix
$$
together with the mass shift
$$
m\to m-\d m\,.
\eqn\massrenorm
$$

It is easy to see that in the self-energy we then
have to order ${\bold v}$ the counterterm:
$$
-i\Sigma_{\bold v}^{\rm count}=i\d Z_2(p\slash-m) +2iZ'(p_0
v\slash - p\cdot v\eta\slash) +i\d m
\,,
\eqn\no
$$
with $Z_2=1+\d Z_2$ and
where we made use of $\eta\slash v\slash=-v\slash\eta\slash$.
We should also recall
that $\delta Z_2$ and $Z'$ are both of order $\alpha$. We see that
a suitable choice of $Z'$ cancels the UV divergences of (\UVs).
Schematically the full self-energy, $-i\Sigma_{\bold v}^{\rm tot}=
-i\Sigma_{\bold v} -i\Sigma_{\bold v}^{\rm count}$, now has the form
$$
-i\Sigma_{\bold v}^{\rm tot}\equiv m\a + p\slash \b + p_0\eta\slash \d
+mv\slash \eps
\,,
\eqn\abcde
$$
where the functions $\a$, $\b$, $\d$ and $\eps$ depend upon $p^2$, $p_0$ and
$p\cdot v$.

The on-shell renormalisation condition is that at the physical mass,
$m$, the dressed propagator, $iS_{\bold v}$, should have the form of a
free field propagator, i.e., with a simple pole at $m$, the Lagrangian
mass, and a residue of one at
that pole. From (\abcde) it then follows that
$$
\tilde\a+\tilde \b +\frac{p_0^2}{m^2}\tilde \d +\frac{p\cdot
v}{m}\tilde \eps\equiv 0
\,,
\eqn\Condition
$$
where the tildes on the functions denote that they are evaluated at
$p^2=m^2$. We stress that the non-covariant nature of these functions
implies that this condition should hold for arbitrary $p_0$ and $p\cdot
v$ as long as the particle is on-shell. We
will return to this in a moment.

With the above notation for the self-energy we can
write the propagator as
$$
\eqalign{
iS_{\bold v}=i\frac{p\slash+m}{p^2-m^2}-\frac1{p^2-m^2}\Big\{\Big.
(2m^2\tilde \Delta & +\tilde\b)p\slash \cr
&+(2m^2\tilde\Delta +\tilde \a +2\tilde \b)m -
p_0\eta\slash\tilde \d-m v\slash\tilde\eps\Big.\Big\}
\,,}
\eqn\curly
$$
where
$$
\tilde \Delta(p_0, p\cdot v) =
\Big(\frac{\pa\a}{\pa p^2} + \frac{\pa\b}{\pa p^2}
+\frac{p_0^2}{m^2}\frac{\pa\d}{\pa p^2} +\frac{p\cdot v}{m}
\frac{\pa\eps}{\pa p^2}\Big)\Bigm|_{p^2=m^2}
\,.
\eqn\XZX
$$
We see that if the second term is to vanish we must require
that the three-vector component of $p$ is proportional to ${\bold v}$.
Let us now take
$$
p^{\mu}=(m,\,m{\bold  v})
\,,
\eqn\chosen
$$
which is what we could expect from our introductory considerations:
$p^2=m^2+O(v^2)$ and the three-momentum is that of a particle with
velocity ${\bold v}$. Our prediction is that this will be a good
on-shell renormalisation point for this dressed propagator.
It then follows that we require for the term in the curly brackets to
vanish on-shell that
$$
\bar\a+\bar\b+\bar\d=0
\,,
\eqn\condone
$$
which is just (\Condition) with the choice (\chosen),
and
$$
\bar \d=\bar\eps
\,
\eqn\condtwo
$$
where bars over the functions mean that they
are now evaluated at $p^\mu=(m,\,m{\bold v})$, our physically motivated
choice of on-shell renormalisation point. Finally we also have
$$
2m^2\bar\Delta+\bar\b -\bar\d=0
\,.
\eqn\condthree
$$

We now want to show that these conditions are indeed satisfied by
the explicit expressions for the propagator from Fig.\ 1 if the
counterterms are chosen properly.
Using the integrals of Ref.\ \ADKINS\ one finds after some work
the following expressions for $\a$, $\b$, $\d$ and $\eps\,$:
$$
\eqalign{
\tilde\a=& i\frac{\d m}m +\frac{i\a}{4\pi}\left[-\frac4{\hat\eps}
-10-2\frac{p_0}{{\bf
p}}\chi  \right]-
i\d Z_2-\frac{i\a}{4\pi}\frac{2p\cdot v}{{\bf p}^3} \left[
2p_0{\bf p}+m^2\chi   \right]
\,,\cr
\tilde\b=& i\frac{\a}{4\pi}\left[\frac1{\hat\eps}
+6-\frac{4p_0^2}{{\bf p}^2} +\(\frac{2p_0}{{\bf p}}-
\frac{2p_0^3}{{\bf p}^3}  \)\chi \right] \cr
&\qquad+ i\d Z_2 +i\frac{\a}{4\pi}\frac{2p\cdot v}{{\bf
p}^5} \left[ 6{\bf p}p_0^3+ 2 {\bf p}^3 p_0 +3 m^2(p_0^2+{\bf
p}^2)\chi\right]
\,,\cr
\tilde\d=& i\frac\a{4\pi}\left[ -4+\frac{4p_0^2}{{\bf p}^2} -
\( \frac{2p_0}{{\bf p}}-\frac{2p_0^3}{{\bf p}^3}\) \chi \right]
-2iZ'\frac{p\cdot v}{p_0}\cr
&-i\frac{\a}{4\pi}\frac{2p\cdot v}{9p_0{\bf p}^5} \left[
\frac{12}{\hat\eps}{\bf p}^5 + 58{\bf p}^5 -24{\bf p}^3p_0^2 +18{\bf
p}p_0^4 +(9p_0^5-15{\bf p}^2p_0^3 +18{\bf p}^4 p_0) \chi \right]
\,,\cr
\tilde\eps=& 2iZ'\frac{p_0}m +i\frac\a{4\pi}\frac{2p_0}{9m{\bf p}^3}
\left[ \frac{12}{\hat \eps}{\bf p}^3 + 46{\bf p}^3 +6 {\bf p} p_0^2
+(3p_0^3+ 9 {\bf p}^2 p_0) \chi \right]
\,,
}
\eqn\expressions
$$
where
$$
{\bf p}=\vert{\bold p}\vert\,,\quad \chi={\rm ln}\(\frac{p_0-{\bf p}}
{p_0+{\bf p}}\)\,,\quad{\rm and}\qquad \frac1{\hat \eps}
=\frac1{2-\omega}-\gamma_E+{\rm ln}4\pi
\,.
\eqn\no
$$
In the limit ${\bold v}\to 0$ we so obtain the results for the
dressed static propagator (note that there are no $1/{\bf p}$
singularities).

One sees that the condition (\Condition) is satisfied
for arbitrary
values of $p_0$ and $p\cdot v$ only if
$$
\d m=\frac{\alpha}{4\pi}(\frac3{\hat\eps}+4)m
\,.
\eqn\no
$$
This may be recognised as exactly the standard
mass shift for the fermion propagator which is
known\up{\DMONE,\BROADHURST,\TOM} to be gauge
parameter independent in the class of Lorentz gauges and
was also found for the
static, i.e., ${\bold v}\to0$, dressed propagator\up{\MONSTER}.

We now have at the physical mass shell
$$
\eqalign{
\bar\a=& i\frac{\d m}m +i\frac\a{4\pi}\left[-\frac4{\hat\eps} -6
\right] - i\d Z_2
\,,\cr
\bar\b=& i\frac\a{4\pi}\left[\frac1{\hat\eps} +\frac{10}3\right]
+i\d Z_2
\,, \cr
\bar\d=& i\frac\a{4\pi}\left[-\frac43\right]
\,, \cr
\bar\eps=& i\frac\a{4\pi}\left[ \frac8{3\hat\eps} +\frac{52}9\right]
+2iZ'
\,,
}
\eqn\no
$$
which means that (\condone) is automatically satisfied --- as we would
expect because of its close relation to (\Condition). Eq.\ \condtwo\
now requires that
$$
Z'=-\frac{\alpha}{4\pi}\(\frac4{3\hat\eps}+\frac{32}9\)
\,.
\eqn\no
$$

To obtain $\bar\Delta$ at order ${\bold v}$ we just
need $\overline{\pa\a/\pa p^2}$ and similar for
$\b$ and $\d$ (the analogous derivative of
$\eps$ does not contribute at this order in
${\bold v}$. One finds
$$
\eqalign{
\overline{\frac{\pa\a}{\pa p^2}}=&\frac{i\a}{4\pi}\frac1{2m^2}
\left[2{\rm ln}\frac{\lambda^2}{m^2}+\frac43\right]\,,\cr
\overline{\frac{\pa\b}{\pa p^2}}=& \frac{i\a}{4\pi}\frac1{2m^2}
\left[ -\frac23{\rm\ln}
\frac{\lambda^2}{m^2}-\frac{122}{45}\right]\,,\cr
\overline{\frac{\pa\d}{\pa p^2}}=&
\frac{i\a}{4\pi}\frac1{2m^2}\left[ -\frac43{\rm ln}
\frac{\lambda^2}{m^2} -\frac{148}{45}\right]\,.
}
\eqn\no
$$
where we have introduced a photon mass $\lambda$ to regulate the
infra-red sector.
With these one has that the various infra-red divergences cancel and
$$
\bar\Delta=\frac{i\alpha}{4\pi}\frac{1}{2m^2}
\left[-\frac{14}3\right]
\,.
\eqn\no
$$
This means that the condition (\condthree) can be reexpressed as
$$
Z_2=1-\frac{\a}{4\pi} \frac1{\hat\eps}
\,.
\eqn\no
$$
This we recognise as the static wave function renormalisation constant.
This concludes the renormalisation of the propagator
to first order in the velocity. A finite result has been obtained and,
up to the need for the matrix rotation parameterised by $Z'$, we have
been able to use completely standard techniques.
This fully supports the conjecture in Ref.\thinspace\MONSTER.

Another consequence of the results of this paper is that we have
introduced a class of gauges parameterised by the vector, ${\bold v}$,
which are infra-red finite. As well as these gauges we are only aware
of the Yennie\up{\YENNIE, \NAKA} gauge and the Coulomb gauge as yielding
infra-red finite propagators. It should be considered
that our propagator depends
upon two vectors, $\eta$ and ${\bold v}$, and we recall the many
difficulties with axial gauge calculations\up{\AXIAL}.

The next
extensions of this calculation to other QED Green's functions are
rather clear: studies for relativistic velocities together with
dressed vertices and higher
Green's functions need to be considered. We hope that explicit results
will further clarify the principles
underlying the infra-red finiteness of these calculations.

The non-abelian extension of this work is still lacking, although the
one-loop dressed quark propagator, whether static or not, is just
its QED equivalent modulo a colour factor at this
order in the coupling. Higher loop studies need to be performed.
Asymptotic freedom means that at short distances perturbative
dressings, like those of QED, are sufficient. At larger distances and
higher orders in the coupling differences appear and the gluonic
dressing around a quark becomes more and more
complicated\up{\MONSTER,\PANP}. Non-perturbatively it is impossible to
dress the quarks\up{\PANP,\MONSTER} and asymptotic quark states cannot
be constructed. This does not mean that the perturbatively dressed
fields do not have a physical significance and we would for example
suggest that their use could be of interest in the heavy quark
effective theory.

\bigskip
In summary we have studied the one loop propagator of a fermion dressed
in such a way as to correspond to a charge moving with a small velocity
${\bold v}$. It was shown that the renormalisation procedure singled
out the expected renormalisation point, $p=(m, m{\bold v})$ in accord
with previous conjectures\up{\MONSTER}. We interpret this as strong
evidence that the dressed fermion (\nonstatic) is a good asymptotic
field. The renormalisation of the propagator at higher orders in the
velocity and indeed a general verification of our conjectures for the
propagator of the dressed field (\nonstatic)
will be presented elsewhere.

\bigskip
\ni {\bf Acknowledgements:} EB thanks the HE group at BNL for its warm
hospitality and DGICYT for financial support.
MJL thanks project CICYT-AEN95-0815 for
support and the TH division at CERN for their hospitality.
\bigskip\bigskip

  \immediate\closeout\reffile
  \centerline{{\bf References}}\bigskip\eightpoint\frenchspacing%
  \input refs.tmp\vfill\eject\nonfrenchspacing

\bye